# Roles of Fe-ion irradiation on MgB$_2$ thin films: Structural, superconducting, and optical properties

Dzung T. Tran [a,*], Tien Le [a,*], Yu-Seong Seo [a], Duc H. Tran [b], Tuson Park [c], Soon-Gil Jung [d], T. Miyanaga [e], Chorong Kim [f], Sunmog Yeo [f], Won Nam Kang [a,**], Jungseek Hwang [a,**]

a. Department of Physics, Sungkyunkwan University, Suwon, 16419, Republic of Korea
b. Faculty of Physics, University of Science, Vietnam National University, Hanoi, 100000, Viet Nam
c. Center for Quantum Materials and Superconductivity (CQMS), Department of Physics, Sungkyunkwan University, Suwon 16419, Republic of Korea
d. Department of Physics Education, Sunchon National University, Suncheon 57922, Republic of Korea
e. Department of Mathematics and Physics, Hirosaki University, Hirosaki, Aomori, 036-8561, Japan
f. Korea Multi-Purpose Accelerator Complex, Korea Atomic Energy Research Institute, Gyeongju, Gyeongbuk 38180, Republic of Korea

*These authors contributed equally to this work*

**Corresponding authors*

*Email addresses:* wnkang@skku.edu (W. N. Kang), jungseek@skku.edu (J. Hwang)

## Abstract

The effects of Fe-ion irradiation on the crystal structure and superconducting properties of MgB$_2$ thin films were investigated. Pristine samples were prepared using hybrid physical-chemical vapor deposition (HPCVD), and ion irradiation was performed at three different doses of 5 x 10$^{13}$, 1 x 10$^{14}$, and 2 x 10$^{14}$ ions/cm$^2$. The measured temperature-dependent resistivity showed that as the irradiation dose increased from pristine to most irradiated, the superconducting critical temperature, $T_c$, significantly decreased from 38.33 to 3.02 K. The crystal structures of the films were investigated by X-ray diffraction (XRD) and X-ray absorption spectroscopy (XAS) measurements. The results showed that the higher the dose, the greater the change in crystal structure, such as the lattice constant and bond length. This suggests that the destruction of the crystal structure at higher doses leads to the degradation of superconductivity in the irradiated MgB$_2$ thin films. Raman spectroscopy showed that the electron-phonon coupling constant decreased with increasing irradiation dose, which was directly related to the reduction of $T_c$ in the samples. The optical conductivity indicates that the charge-carrier density of the $\sigma$-band plays an important role in the superconductivity of ion-irradiated MgB$_2$.

**Introduction**

Among metallic superconductors, magnesium diboride has many advantages such as a high superconducting critical temperature ($T_c \sim 39$ K), a high critical current density ($J_c$) of approximately $10^7$ A/cm$^2$, a simple crystal structure compared with other high-temperature superconductors (HTSs), and a low-cost growth process that does not require rare-earth elements. However, because the crystal structure of MgB$_2$ thin films is nearly perfect, the lack of natural vortex pinning centers in MgB$_2$ causes a rapid decrease of $J_c$ by the applied magnetic field. Many studies on ion-irradiated MgB$_2$ thin films have shown that ion irradiation is an effective method for creating defects in the crystal structure and improving $J_c$. Therefore, a complete understanding of the effects of ion-irradiation on the MgB$_2$ matrix is crucial. Tarantini *et al.* irradiated 1.47 and 0.84 MeV neutrons on MgB$_2$ samples [1]. They observed that after irradiation, the volume of the unit cells increased by 1.7%. This expansion created disorder in the crystal structure and indirectly created defects. The $T_c$ of the irradiated sample decreased to 9 K at a dose of 1.4 x $10^{20}$ cm$^{-2}$. Wilke *et al.* reported a different result for neutron irradiation: the cell volume increased by 2.6% at a dose of 9.5 x $10^{18}$ cm$^{-2}$ with a $T_c$ of approximately 37 K [2]. Le *et al.* recently used 2 MeV Sn ion irradiation on the thin MgB$_2$ films. The unit cell volume increased by approximately 9% compared to that of the pure MgB$_2$ samples. The degradation of $T_c$ was also observed, with a $T_c$ of 29.6 K [3]. On the contrary, there was a study on doping Fe nanoparticles into both MgB$_2$ bulk and thin films, showing that the lattice constants remain unchanged in the doped samples, however both $T_c$ and $J_c$ experienced a declining trend because of the presence of Fe nanoparticles in the matrix of the materials [4]. Up until now, many studies have been performed to improve the $J_c$ and $T_c$ using ion and neutron irradiations, as discussed above. However, Fe-ion irradiation effects on MgB$_2$ thin films have not been fully studied yet. Therefore, we focused on the physical properties of Fe ion-irradiated MgB$_2$ samples, including optical properties, and explained the superconducting properties of the irradiated MgB$_2$ using the experimental observations.

In this study, we investigated the effects of Fe ion irradiation at an energy of 140 keV on MgB$_2$ thin films. The measured $T_c$ and the change in the upper critical field ($\mu_0 H_{c2}$) were discussed. X-ray diffraction (XRD), X-ray absorption spectroscopy (XAS), and Raman spectroscopy were used to investigate the crystal structure and fully explain the mechanism of changing $T_c$ of irradiated samples. The electron-phonon coupling constant was estimated from the measured Raman spectra using the McMillan equation with an Allen-Dynes modification. The displacement per atom (*dpa*) relation was used to describe the effect of disorder on the

MgB$_2$ thin films [5]. Finally, optical conductivity spectra were obtained from the reflectance spectra measured using a Fourier transform infrared (FTIR) spectrometer. The evolution of charge carrier density caused by Fe-ion irradiation was examined from the optical conductivity using the Drude-Smith model analysis. The optical results indicate that the charge carrier density of the $\sigma$-band is closely related to $T_c$ in the Fe-ion-irradiated samples.

**Experiment**

Using hybrid physical-chemical vapor deposition (HPCVD), MgB$_2$ thin films with a thickness of 192 nm were fabricated on *c*-cut Al$_2$O$_3$ substrates. Details of the fabrication process are described elsewhere [6-9]. Pristine samples were irradiated with an Fe-ion beam with an energy of 140 keV at three different doses of 5 x 10$^{13}$, 1 x 10$^{14}$, and 2 x 10$^{14}$ ions/cm$^2$ at the Korea Multipurpose Accelerator Complex (KOMAC) at 300 K. The mean projected range ($R_p$) of Fe ions in MgB$_2$ films was estimated using the stopping and range of ions in matter (SRIM) program based on the Monte Carlo equation [10, 11]. The crystalline structures of the pristine and ion-irradiated samples were investigated by XRD using a Miniflex 600 instrument. A four-probe technique was used to measure the DC resistivity as a function of temperature under magnetic fields applied along the *c*-axis of the samples. Resistivity measurements were performed using a physical property measurement system (PPMS) (9 T, Quantum Design). To examine the local structure of the MgB$_2$ samples, extended X-ray absorption spectroscopy (EXAFS) was performed using a photon factory beamline spectrometer (BL11A) of the High Energy Accelerator Research Organization (KEK), Tsukuba, Japan. The incident X-ray beam was parallel to the *ab*-plane of the MgB$_2$ thin film. The Mg K-edge EXAFS spectrum was plotted in the electron yield mode and fitted using ATHENA and ARTEMIS in the Demeter Perl software code of IFEFFIT. Raman spectroscopy was performed at 300 K using a Raman spectrometer (MAPLE II) with a 532 nm green laser as the incident beam. Several Raman spectroscopy measurements were performed for each sample to obtain the average Raman signals. The Raman shift ranged from 200 to 1400 cm$^{-1}$ in an exposure time of 10 s. The reflectance spectra ($R(\omega)$) of the samples in a wide spectral range (170–22000 cm$^{-1}$) were measured using a commercial FTIR spectrometer (Bruker Vertex 80v). The optical conductivity was obtained from the measured reflectance using the Kramers-Kronig analysis [12]. To perform the Kramers-Kronig analysis, the measured reflectance spectrum must be extrapolated to both zero and infinite frequencies. For the extrapolation to zero, the Hagen-Rubens relation $(1 - R(\omega) \propto \sqrt{\omega})$ was used. For the extrapolation to infinity,

$R(\omega) \propto \omega^{-1}$ was used from 22000 to $10^6$ cm$^{-1}$ and above $10^6$ cm$^{-1}$, free electron behavior ($R(\omega) \propto \omega^{-4}$) was assumed.

**Results and discussions**

The appropriate type and energy of ions for this study were determined by simulation using the SRIM software. Fig. 1 shows the mean projected range of a 140 keV Fe ion beam irradiating MgB$_2$ as simulated by the SRIM software. A 192 nm thick MgB$_2$ film with a density of 2.57 g/cm$^3$ on a sapphire substrate was set as the target material for the simulation. The depth profile shows that the Fe ions are typically concentrated in the MgB$_2$ layer with a mean projected range ($R_p$) of 96 nm without damaging the Al$_2$O$_3$ substrate. The Fe ions in the MgB$_2$ film layer may act as Frenkel point defects and distort the MgB$_2$ crystal structure, forming non-superconducting regions nearby, alongside the superconducting region [13]. Note that the energy and mass of the Fe ion are not enough to introduce columnar defects in the MgB$_2$ film layer [14]. The number of vacancies in the sample after irradiation suggests that both the ion energy and type are affected by the displacements in the structure. Therefore, the displacement per atom (*dpa*) was estimated using the following equation [15]:

$$dpa = \frac{\text{number of vacancies}}{\text{Å} \times \text{ion}} \times \left[\frac{10^8 \text{ Å/cm}}{\text{atomic density}} \times \text{dose}\left(\frac{\text{ions}}{\text{cm}^2}\right)\right] = \frac{\text{number of vacancies}}{\text{atoms}},$$

where the atomic density of MgB$_2$ is $10.11 \times 10^{22}$ atom/cm$^3$. The average *dpa* (*dpa*$_{\text{avg}}$) was calculated as follows: $dpa_{\text{avg}} = \frac{dpa}{\text{sample thickness}}$. The *dpa*$_{\text{avg}}$ values for the doses of 0, 5x10$^{13}$, 1x10$^{14}$, and 2x10$^{14}$ ions/cm$^2$ are 0, 0.028, 0.057, and 0.113, respectively.

The superconducting critical temperatures (*T*$_c$) of all samples were obtained from the measured DC resistivity data at zero magnetic field, as shown in Fig. 2. As the Fe ion irradiation dose increased, the measured DC resistivity (*ρ*) of the samples increased through the entire temperature range. The *T*$_c$ of the pristine sample was approximately 38.33 K. The *T*$_c$ of the pristine sample was not exactly 39 K because of the thickness of the thin film; the thin film was not thick enough for grain coalescence on the thin film surface [16]. The *T*$_c$ of the samples decreased with irradiation and the *T*$_c$ of the most irradiated sample with *dpa*$_{\text{avg}}$ = 0.113 was approximately 3.02 K. This suggests that the Fe ions embedded in the sample destroyed the crystal structure of the MgB$_2$ films. Furthermore, magnetic fields of 1 T, 3 T, 5 T, 7 T, and 9 T were applied along the *c* axis of each sample to obtain the upper critical field (*μ*$_0$*H*$_{c2}$) of pristine and irradiated samples. Fig. 3 (a) shows the normalized DC resistivity (*ρ(T)*/*ρ*(40 K)) at various magnetic fields. Notably, the *dpa*$_{\text{avg}}$ = 0.113 sample completely loses its superconductivity at the applied magnetic field above 1 T; therefore, we could not obtain the $\mu_0H_{c2}$ for this sample. The temperature where the horizontal line intersects with the DC resistivity represents the $T_c$ used to determine $\mu_0H_{c2}$. The determined $\mu_0H_{c2}$ as functions of temperature are shown in Fig. 3 (b). The two-band Ginzburg-Landau theory was used to fit the $\mu_0H_{c2}$ curves. The fitting model for $\mu_0H_{c2}$ at $dpa_{avg}$ = 0, 0.028, and 0.057 are as follows:

$$\mu_0H_{c2} = \frac{\theta^{1+\alpha}}{1-(1+\alpha)\omega+l\omega^2+m\omega^3},$$

where $\omega = (1-\theta)\theta^{1+\alpha}$, $\theta = 1 - T/T_c$, $\alpha$, $l$, and $m$ are the fitting parameters [17, 18]. The best fits (solid lines) of the experimental data are shown in Fig. 3(b). Interestingly, the upper critical field (10.6 T) of the $dpa_{avg}$ = 0.028 sample at 0 K was higher than that of the pristine sample (8.5 T). Further increase in the irradiation dose deteriorated the $\mu_0H_{c2}$; the $\mu_0H_{c2}$ of the $dpa_{avg}$ = 0.057 sample is approximately 7 T. It is because the irradiated Fe ions act as defects and create disorder regions in the sample, which play the role of flux-pinning centers, pinning down the vortices movement. The results show that the flux-pinning centers enhance the $\mu_0H_{c2}$ up to the optimal dose. However, above the optimal dose, Fe ions and defects also suppress superconductivity too much, resulting in a decrease in $\mu_0H_{c2}$. A similar result was reported in a study on oxygen ion-irradiated $MgB_2$ thin films [19].

To study the crystal structures of the pristine and irradiated $MgB_2$ films, XRD measurements were performed with the $2\theta$ scan mode from 20 to 70 degrees. The XRD data for all samples are shown in Fig. 4. The sharp peaks at around 25° and 52° correspond to the (0001) and (0002) planes, respectively, indicating that the samples were grown with a high c-axis orientation. The full width at half maximum increased from approximately 0.29° for the pristine sample to 0.31° for the irradiated sample at $dpa_{avg}$ = 0.113, indicating that the crystallinity deteriorated with increasing irradiation dose. The (0001) and (0002) peaks of the irradiated samples were slightly shifted to lower angles compared to those of the pristine sample, implying that the lattice parameter of the irradiated $MgB_2$ films increased. The (0002) peaks for all samples exhibited a small shoulder on the higher-angle side, indicating the existence of a small amount of $Mg_{1-x}Al_xB_2$ [9, 20]. The c-axis lattice constants of all samples were determined from the corresponding (0002) peaks using Bragg's law, as shown in Table 1. Compared with the pristine sample, a systematic expansion of the c-axis lattice constant was observed from approximately 3.52 to 3.58 Å with increasing the irradiation dose. The Fe-ion irradiation causes two effects: the lattice distortions caused primarily by the collisions between the Fe ions and Mg or B atoms, which create defects and vacancies that assist vortex pinning and consequently increase the $\mu_0H_{c2}$, and the destruction of the crystal structure and consequently reducing the $T_c$. It has been suggested that the irradiation introduced disorders of lattice structure that might result in both intraband and interband scatterings between $\sigma$- and $\pi$-bands, and the interband scattering led to a reduction of $T_c$ [19].

The crystal structure of MgB$_2$ is anisotropic and hexagonal, with the *a*- and *c*-axis lattice constants representing the in- and out-of-plane Mg-Mg bond distances, respectively. Because the XRD data show only the (0001) and (0002) peaks, they only provide information on the out-of-plane Mg-Mg bond. Therefore, to study the distortions in the local structure of MgB$_2$, Mg K-edge EXAFS spectra ($k\chi(k)$) were measured and analyzed, where $k$ is the photoelectron wavenumber, which can be described by $k = \sqrt{2m_e(E - E_0)}/\hbar$, where $m_e$ is the electron mass, $E$ is the x-ray energy, and $E_0$ is the energy of the absorption edge. The measurements were performed parallel to the surface of the MgB$_2$ thin films. The Fourier transformations (FT($k\chi(k)$)) of $k\chi(k)$ of the pristine and irradiated samples are shown in Fig. 5 (a). Two main peaks marked at approximately 1.7 and 2.5 Å match up with the reported Mg-B and Mg-Mg scatterings, respectively [3, 21]. Compared to the pristine sample, the two main peaks of all irradiated samples shifted to higher radial distances, suggesting changes in the local structure of the MgB$_2$ thin films. To quantitatively investigate the local atomic displacement, a double shell model for the Mg-B bond and in-plane Mg-Mg bonds was applied to fit the FT Mg K-edge EXAFS spectra. Throughout the fitting process, all the parameters were fixed except for the bond distances ($R$) and the mean-square relative displacement ($\sigma^2$) known as the Debye-Waller factor. The fitting ranges of $k$ and $R$ were from 3 to 8 Å$^{-1}$ and from 1.3 to 3 Å, respectively. The data and fit for the pristine sample are shown in the inset of Fig. 5 (a). The $R$ and $\sigma^2$ of Mg-B and Mg-Mg bonds of all samples obtained from the fitting as functions of $dpa_{avg}$ are shown in Fig. 5(b). The bond length of Mg-B increased as the irradiation dose increased, confirming the expansion of the *c*-axis lattice constant. This expansion interrupts the movement of electrons between the Mg and B layers, resulting in the suppression of $T_c$. In addition, the in-plane Mg-Mg bond length was practically unchanged, indicating that the lattice constant along the *a*-axis was not affected by the irradiation process. The Debye-Waller factor shows a similar irradiation dose-dependent trend as the bond distance, which represents the correlation between the local atomic disorder and bond length. Our results agree with those of other studies [3]. However, a previous study of Fe-doped MgB$_2$ was conducted on both bulk and thin-film MgB$_2$ [4]. The results showed that even though the Fe substitution for Mg in the lattice led to suppression for both $T_c$ and $J_c$ the lattice constants remained unchanged in the doped samples, which is a different behavior from what we observed in this study.

Defects and/or lattice disorders may reduce the superconductivity of MgB$_2$. Raman spectroscopy can be used to study the lattice disorders or defects caused by Fe-ion irradiation. The Raman spectra of the pristine and Fe-ion-irradiated MgB$_2$ samples are shown in Fig. 6. For the pristine MgB$_2$ crystal, only one Raman-active soft mode ($E_{2g}$) is associated with the vibrations of in-plane boron atoms [22, 23]. The $E_{2g}$ peak is centered at approximately 600 cm$^{-1}$ and assigned as $\omega_2$ in the figure. However, this peak can change because of defects and/or lattice distortions [19, 24]. For getting an accurate fit to the measured Raman spectrum of the pristine sample, two very weak additional peaks are required, which are assigned as $\omega_1$ and $\omega_3$ centered at approximately 400 and 780 cm$^{-1}$, respectively. In addition, comparing the intensities of $\omega_2$ and $\omega_3$ peaks of the pristine and irradiated samples, an opposite trend was observed between these two peaks; the intensity of the $\omega_3$ peak significantly increases, while that of the $\omega_2$ peak decreases, making the overall Raman spectra shift toward the higher frequency. The high-frequency structure at approximately 780 cm$^{-1}$ was assigned as the peak in the phonon density of states (PDOS) [25-27], which became dominant after Fe-ion irradiation and was caused by the defects introduced by the irradiation, which increased the Mg-B bond length and distorted the boron plane. We speculate that these distortions in the boron plane influence the vibration of boron atoms and may result in the suppression of the $E_{2g}$ mode. Similar features have been previously reported [19,28]; the high-frequency spectral structures centered at approximately 700 and 790 cm$^{-1}$ were observed in the Raman spectra of an $O^{2+}$-ion-irradiated MgB$_2$ [19] and neutron irradiated MgB$_2$ samples [28]. Wang *et al.* [19] claimed that the shift in the Raman spectra was due to a change in the volume of the crystal unit cell caused by the ion irradiation, and the relaxation of the Raman selection rules in the system that has defects could lead to the detectable rise of high-frequency regions. Note that, comparing with the results of Wang *et al.*'s study, our lowest irradiation dose is too high to show the intermediate state, including the high-frequency feature near 700 cm$^{-1}$.

The $T_c$ of a superconducting system can be expressed as a function of the average phonon frequency using the McMillan formula with an Allen-Dynes modification [5] as follows:

$$T_c = \frac{\langle \omega \rangle}{1.2} \exp\left(-\frac{1.04(1+\lambda)}{\lambda - \mu^*(1+0.62\lambda)}\right),$$

where $\langle\omega\rangle$ ($\equiv (\omega_1 + \omega_2^2 + \omega_3)^{0.25}$) is the average phonon frequency, $\mu^* = 0.12$ is the Coulomb pseudopotential for MgB$_2$ [29], and $\lambda$ represents the electron-phonon coupling constant. The Debye temperature can also be calculated from the average phonon frequency using the following equation:

$$\Theta_\mathrm{D} = \frac{\langle\omega\rangle}{1.2} \times 1.45.$$

Table 1 shows the electron-phonon coupling constant ($\lambda$) and Debye temperature ($\Theta_\mathrm{D}$) of all samples. The $\lambda$ systematically decreases as the irradiation dose increases, from 1.03 for the pristine sample to 0.45 for the most irradiated one, which might be directly related to the reduction of $T_\mathrm{c}$. According to the Bardeen-Cooper-Schrieffer (BCS) theory, two electrons attract each other by exchanging a force-mediated phonon and form an electron-electron Cooper pair. Therefore, the weakened electron-phonon coupling results in weakly formed Cooper pairs in the superconductor, resulting in a lower $T_\mathrm{c}$. MgB$_2$ is known as a two-band superconductor and exhibits two superconducting (SC) gaps: the large $\sigma$-band gap formed by the boron $p_\mathrm{x}$ and $p_\mathrm{y}$ orbitals, and small $\pi$-band gap formed by the boron $p_\mathrm{z}$ orbital [30]. The electrons in the $\sigma$-band are strongly coupled to phonons confined in the honeycomb boron layer and give rise to a large SC gap, whereas a relatively small SC gap appears in the $\pi$-band owing to the weak electron-phonon coupling [31]. Therefore, the superconducting state in MgB$_2$ is primarily formed from the $\sigma$-band of the B layers [32]. Mazin *et al.* calculated the coupling constant for the two-band system; the electron-phonon coupling constant for the $\sigma$-band ($\lambda_{\sigma\sigma}$) is 1.02, and that for the $\pi$-band ($\lambda_{\pi\pi}$) is 0.45 [33]. Because the most irradiated sample shows a small $\lambda$ value of 0.45 and a very low $T_\mathrm{c}$ of 3.02 K, the irradiation process may selectively suppress the $\lambda_{\sigma\sigma}$, supporting the fact that the $\sigma$-band plays an important role for the superconductivity in MgB$_2$. The disorder and defects introduced by irradiation weaken the electron-phonon coupling constant and are responsible for the reduction of $T_\mathrm{c}$. However, interestingly, the $\Theta_\mathrm{D}$ abruptly increases from the pristine sample to the irradiated one, and further increases in the irradiation dose have little effect. The $\Theta_\mathrm{D}$ of the pristine sample is 696.82 K, $\Theta_\mathrm{D}$ of the $dpa_\mathrm{avg}$ = 0.028 sample is 739.95 K, and $\Theta_\mathrm{D}$ of the most irradiated ($dpa_\mathrm{avg}$ = 0.113) sample is 745.34 K. These values agree with those previously reported by Susner *et al.*, whose values were approximately 600–750 K [34].

The reflectivity spectra of our MgB$_2$ series samples were measured and analyzed to study their phononic and electronic structures. The measured reflectivity spectra of our MgB$_2$ series samples below 1500 cm$^{-1}$ at 300 K are shown in Fig. 7. The reflectance of the pristine sample was nearly 100%, with nearly no phonons or other significant features. In contrast, the reflectance spectra of the irradiated samples were significantly suppressed and exhibited additional sharp features, as indicated by the vertical dashed lines. The inset of Fig. 7 shows the reflectance spectra in a wide spectral range from 170 to 22000 cm$^{-1}$. Comparing our results with those of other studies by Seo *et al.* and Barker, Jr., some of the additional features were identified as longitudinal optical (LO) phonon ($E_u$) modes of $Al_2O_3$, indicating that the substrate $Al_2O_3$ was exposed after the irradiation process [35, 36]. We note that LO modes are not typically IR-active. However, for ionic compounds such as $Al_2O_3$, the LO mode may appear in the reflectance spectrum along with the corresponding transverse optical (TO) mode [36]. We also note that LO phonons may be IR-active even for non-ionic compounds when the sample has a miscut surface or surface irregularities [37, 38]. The high reflectance in the low-frequency region is attributed to the large charge carrier density in the $MgB_2$ films. Therefore, the decrease in the reflectance of the irradiated samples can be attributed to an increase in the scattering rate, caused by irradiation-induced defects and distortions in the crystal structure of the samples.

The optical conductivity ($\sigma_1(\omega)$) was obtained from the measured reflectance using a Kramers-Kronig analysis [12]. The optical conductivity spectra of the four samples in a wide spectral range up to 22000 cm$^{-1}$ are shown in Fig. 8(a). The spectral weight seems to be decreasing as the irradiation increases, for which we do not clearly know the reason yet. We speculate that the effective mass of the charge carriers may increase with irradiation. Therefore, we also show their normalized accumulated spectral weights in Fig. 8(b) to show the irradiation-dependent spectral weight transfer. The normalized spectral weight (SW) is defined by $\frac{SW(\omega)}{SW(22000 \text{ cm}^{-1})} = \frac{\int_0^\omega \sigma_1(\omega')d\omega'}{SW(22000 \text{ cm}^{-1})}$. We clearly observed the spectral weight transfer from low to high energy as the irradiation increased. Figs. 8(c)-(f) show the optical conductivity spectra and fits below 4000 cm$^{-1}$ using Drude and Drude-Smith models. The free charges in the pristine ($dpa_{avg}$ = 0) sample can be described by the Drude model, which has two parameters: the Drude plasma frequency ($\omega_{p,D}$) and impurity scattering rate ($1/\tau_{imp,D}$). In contrast, the charge carriers in the irradiated samples can be described using the Drude-Smith model [12, 39], which has three parameters: the Drude-Smith plasma frequency ($\omega_{p,DS}$), impurity scattering rate ($1/\tau_{imp,DS}$), and coefficient $c_1$. The coefficient $c_1$ is the faction of the electron's original velocity that is retained after the first collision. The plasma frequency square is proportional to the charge carrier density. The optical conductivity spectrum in the low-frequency region was fitted to either a narrow Drude (for the pristine sample) or narrow Drude-Smith (for the irradiated samples) mode and an additional broad Drude mode. The real part of the optical conductivity can be written in a narrow Drude-Smith and a broad Drude model as follows [39-41]:

$$\sigma_1(\omega) = \frac{\omega_{p,DS}^2}{4\pi} \frac{1/\tau_{imp,DS}}{\omega^2 + [1/\tau_{imp,DS}]^2} \times \left\{1 + \frac{c_1 \times \left[\left(\frac{1}{\tau_{imp,DS}}\right)^2 - \omega^2\right]}{\left(\frac{1}{\tau_{imp,DS}}\right)^2 + \omega^2}\right\} + \frac{\omega_{p,D}^2}{4\pi} \frac{1/\tau_{imp,D}}{\omega^2 + [1/\tau_{imp,D}]^2}.$$

Notably, in the case of the pristine ($dpa_{avg} = 0$) sample, the $c_1$ is 0, which makes the Drude-Smith model turn into the Drude model. The narrow Drude and broad Drude modes have been assigned as the $\sigma$-band and $\pi$-band, respectively [35, 42]. The plasma frequencies ($\omega_{p,DS}$ and $\omega_{p,D}$) and impurity scattering rates ($1/\tau_{imp,DS}$ and $1/\tau_{imp,D}$) of both the narrow Drude-Smith mode and broad Drude mode were obtained from the fitting and shown in Table 2. The plasma frequency of the broad Drude mode increased slightly from the pristine sample to the irradiated sample and was nearly unaffected by further irradiation. The narrow Drude mode became the Drude-Smith mode in the irradiated samples. The plasma frequency of the narrow Drude-Smith mode of the irradiated sample was dramatically suppressed compared to that of the Drude mode of the pristine sample and kept decreasing with increasing the irradiation dose, indicating that the $\sigma$-band was significantly affected by the irradiation. Therefore, the $\sigma$-band may play a crucial role in the superconductivity of the irradiated $MgB_2$. The peak at around 500 cm$^{-1}$ in all irradiated samples is associated with localized charge carriers caused by Fe-ion irradiation, which was captured by the Drude-Smith model. Note that the electron localization in the $\sigma$-band may be closely associated with the size of the effective $MgB_2$ grain in the ion-irradiated $MgB_2$ film, based on the fitting parameter $c_1$. The more irradiation, the smaller the grain size, resulting in more electrons being localized. Consequently, the $T_c$ will be reduced.

We can clearly observe five phonon peaks that give rise to the $\sigma_1(\omega)$ spectra of all ion-irradiated samples, as marked with A, B, C, and asterisks (*) in Fig. 8. The three (A, B, and C) phonon peaks were identified as the infrared-active phonon modes in $MgB_2$. The two (A and B) peaks are the infrared-active modes, located at approximately 310 and 400 cm$^{-1}$ and are assigned to the $E_{1u}$ and $A_{2u}$ modes, respectively [22]. The reduction in the narrow Drude plasma frequency in the pristine $MgB_2$ allows these modes to appear [43]. At approximately 620 cm$^{-1}$, there is another broad phonon (C) peak, which is a known Raman-active mode, $E_{2g}$ [22]. This mode appears as an infrared-active mode owing to lattice distortion and defects in the $MgB_2$ thin film caused by ion irradiation [43]. Two peaks near 510 and 940 cm$^{-1}$ marked with asterisks (*) do not originate from the MgB$_2$ film. Seo *et al.* have suggested that the two peaks near 510 and 940 cm$^{-1}$ have originated from the Al$_2$O$_3$ substrate [35]. The appearance of the peaks suggests that the surface of the Al$_2$O$_3$ substrate was exposed after ion irradiation.

**Conclusion**

In summary, we investigated the effects of 140 keV Fe-ion irradiation on the crystal structure and superconductivity of MgB$_2$ thin films using three different doses of irradiation. We estimated the $dpa_{avg}$ values of the three irradiated samples as 0.028, 0.057, and 0.113. The results show that irradiation significantly reduced the $T_c$ of the samples. Intriguingly, the sample with $dpa_{avg}$ = 0.028 showed the highest $\mu_0 H_{c2}$, indicating that this irradiation is near the optimal irradiation dose for the $\mu_0 H_{c2}$. The initial ion irradiation created defects, such as vacancies and disorders in the lattice, which act as vortex pinning centers and consequently improve the $\mu_0 H_{c2}$. Meanwhile, a further increase of the irradiation dose may distort and destroy the crystal structure of MgB$_2$ thin film and lead to the reduction of the $T_c$ resulting in the drastic decrease in $\mu_0 H_{c2}$. While the *c*-lattice constant increased with increasing irradiation dose, the *a*-lattice constant remained unchanged. In addition, the suppression of $T_c$ was observed to be directly related to the reduction of the electron-phonon coupling constant ($\lambda$) caused by the defects and/or lattice distortions caused by the irradiation. The optical study showed that the $\sigma$-band (or the narrow Drude band) significantly affected by the irradiation, which is similar irradiation dose-dependent trend of the $T_c$, indicating that the $\sigma$-band plays an important role for the superconductivity of MgB$_2$.

**Declaration of competing interest**

The authors declare no competing financial interests or personal relationships that may have influenced the results reported in this study.

**Acknowledgments**

This study was supported by the Basic Science Research Program through the NRF of Korea, funded by the Ministry of Education (NRF-2021R1F1A1060776 and NRF-2021R1I1A1A01043885), National Research Foundation of Korea (NRFK Grant No. 2021R1A2C101109811), and BrainLink program funded by the Ministry of Science and ICT through the National Research Foundation of Korea (2022H1D3A3A01077468). Miyanaga acknowledges the synchrotron radiation experiments performed at the Photon Factory in KEK under Proposal 2022G050. We wish to acknowledge the outstanding support of the accelerator group and operators of KOMAC and KAERI.

**Figure and Table Captions**

Fig. 1: Ion ranges of irradiated 140 keV iron (Fe) ions into a 192 nm-thick MgB$_2$ thin film on the *c*-cut Al$_2$O$_3$ substrate. The mean project range ($R_\text{p}$) is approximately 96 nm. Number of vacancies created by 140 keV Fe-ion irradiation as a function of target depth. The concentration of Fe atom, ion ranges, and target vacancies were simulated using the SRIM program.

Fig. 2: Temperature-dependent DC resistivity data of the MgB$_2$ thin films with the $dpa_\text{avg}$ values of 0, 0.028, 0.057, and 0.113.

Fig. 3: (a) Temperature-dependent normalized resistivity, $\rho(T)/\rho(40\text{ K})$ of MgB$_2$ thin films with the $dpa_\text{avg}$ values of 0, 0.028, 0.057, and 0.113 at various applied magnetic fields of 1, 3, 5, 7, and 9 T normal, respectively, to the film surfaces. (b) Upper critical field $\mu_0 H_{c2}$ of the pristine and two irradiated MgB$_2$ films with the $dpa_\text{avg}$ values of 0.028, 0.057.

Fig. 4: X-ray diffraction (XRD) patterns of MgB$_2$ films with the $dpa_\text{avg}$ values of 0, 0.028, 0.057, and 0.113. As the $dpa_\text{avg}$ value increases, the peaks for the (0001) and (0002) planes of MgB$_2$ shift to the lower angle.

Fig. 5: (a) Fourier transforms of $k\chi(k)$ Mg K-edge EXAFS spectra of the pristine and Fe ion-irradiated MgB$_2$ thin films. The inset shows the double shell fit for the pristine ($dpa_\text{avg} = 0$) sample as an example. The fit was obtained using Artemis and the IFEFFIT code. (b) Local bond distances of Mg-B and Mg-Mg bonds and the values of the Debye-Waller factor ($\sigma^2$) as functions of the $dpa_\text{avg}$ value.

Fig. 6: Normalized Raman spectra of MgB$_2$ thin films with the $dpa_\text{avg}$ values of 0, 0.028, 0.057, and 0.113 along with fits using Gaussian functions. The three Gaussians modes are $E_{2g}$ mode ($\omega_2$) and two PDOS modes ($\omega_1$ and $\omega_3$).

Fig. 7: Measured reflectivity spectra of the MgB$_2$ thin films with the $dpa_\text{avg}$ values of 0, 0.028, 0.057, and 0.113 at 300 K. The inset shows the measured reflectance spectra of all four samples in a wide spectral range up to 22000 cm$^{-1}$.

Fig. 8: (a) The real parts of the optical conductivity of the MgB$_2$ thin films with the $dpa_\text{avg}$ values of 0, 0.028, 0.057, and 0.113 at 300 K in a wide spectral range up to 22000 cm$^{-1}$. (b) The normalized spectral weights (SW) of the four samples. (c)-(f) The real parts of the optical spectral conductivity below 4000 cm$^{-1}$ along with the corresponding fits obtained using a Drude-Smith and a broad Drude model. The black solid line and red solid line are the data and fits, respectively. The blue dashed line and green dotted line are the Drude-Smith mode and broad Drude mode, respectively. Note that, for the pristine sample ($dpa_{avg}$ = 0), the Drude-Smith mode is essentially the Drude mode because the parameter $c_1$ is 0. The three Γ-point infrared-active phonon modes are marked with A, B, and C. The two peaks marked with the asterisks (*) are originated from the $Al_2O_3$ substrate.

Table 1: Physical quantities of the pristine and Fe ion-irradiated $MgB_2$ thin films with the $dpa_{avg}$ values of 0, 0.028, 0.057, and 0.113: the $c$-lattice constant, the DC resistivity ($\rho$) at 40 K, the SC critical temperatures ($T_c$), the upper critical field ($\mu_0 H_{c2}$), Raman peak centers ($\omega_1$, $\omega_2$, and $\omega_3$), the electron–phonon coupling constants ($\lambda$) estimated using the McMillan equation with Allen-Dynes modification, and the Debye temperatures ($\Theta_D$).

Table 2: Fitting parameters obtained using a Drude-Smith mode and a broad Drude mode: the plasma frequency, impurity scattering rate, and parameter $c_1$ ($\omega_{p,DS}$, $1/\tau_{imp,DS}$, and $c_1$) of the Drude-Smith mode and the plasma frequency and impurity scattering rate ($\omega_{p,D}$ and $1/\tau_{imp,D}$) of the broad Drude mode.

Fig. 1

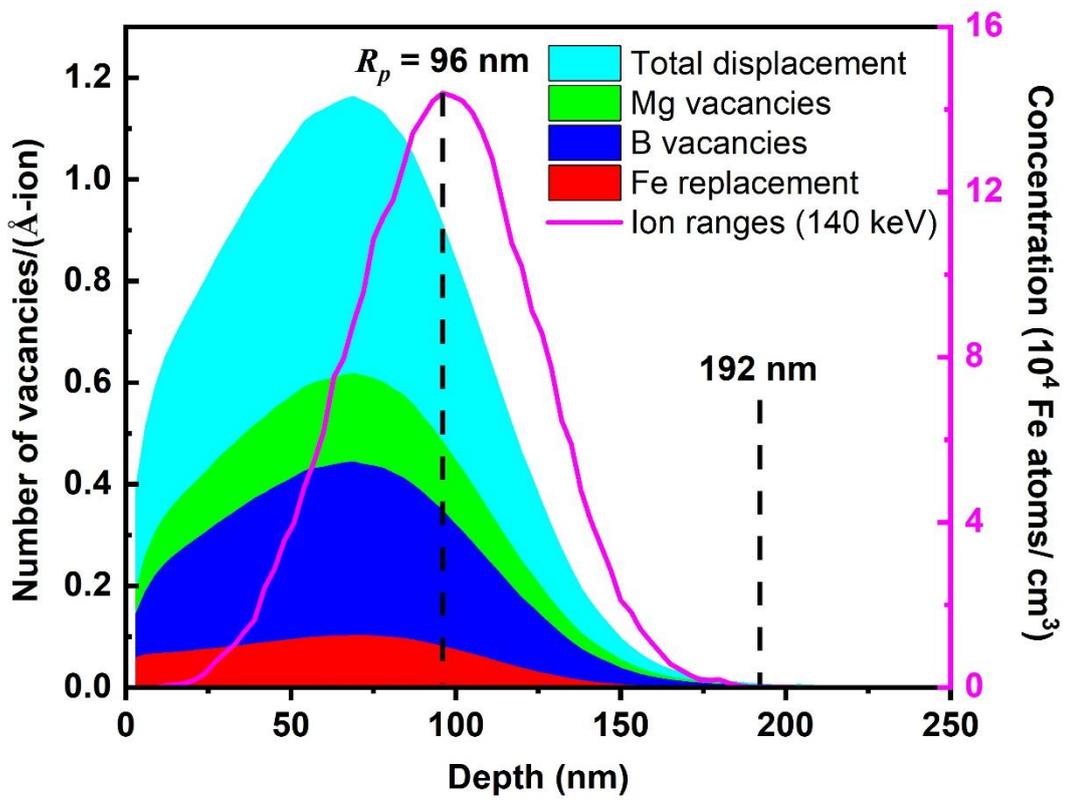

Fig. 2

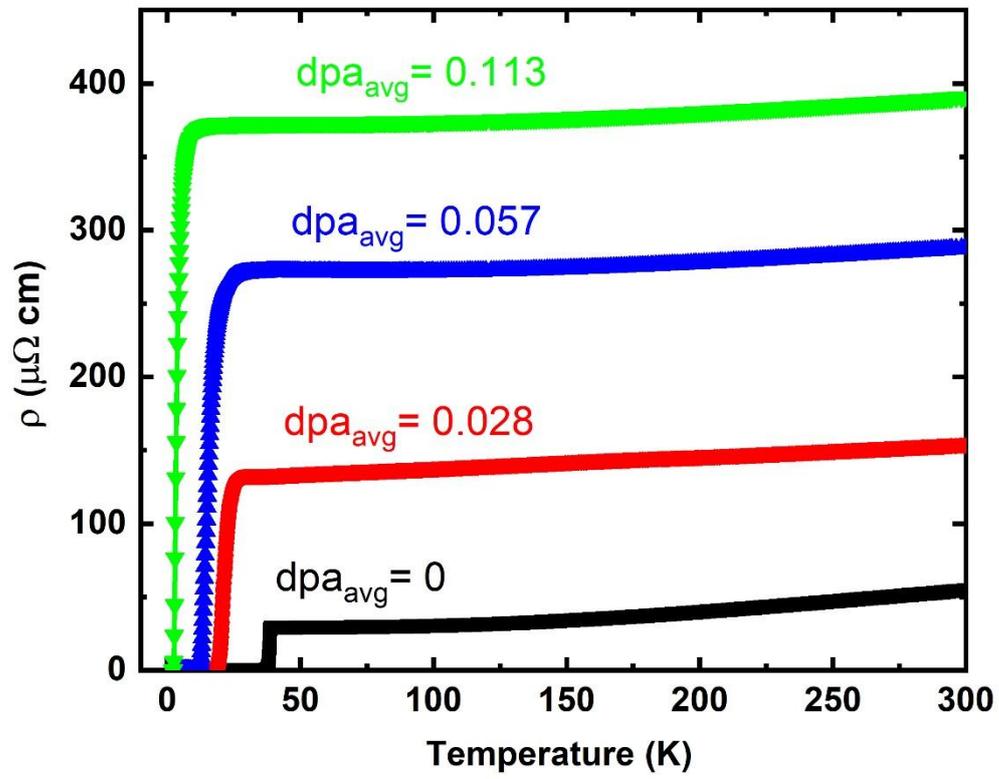

Fig. 3

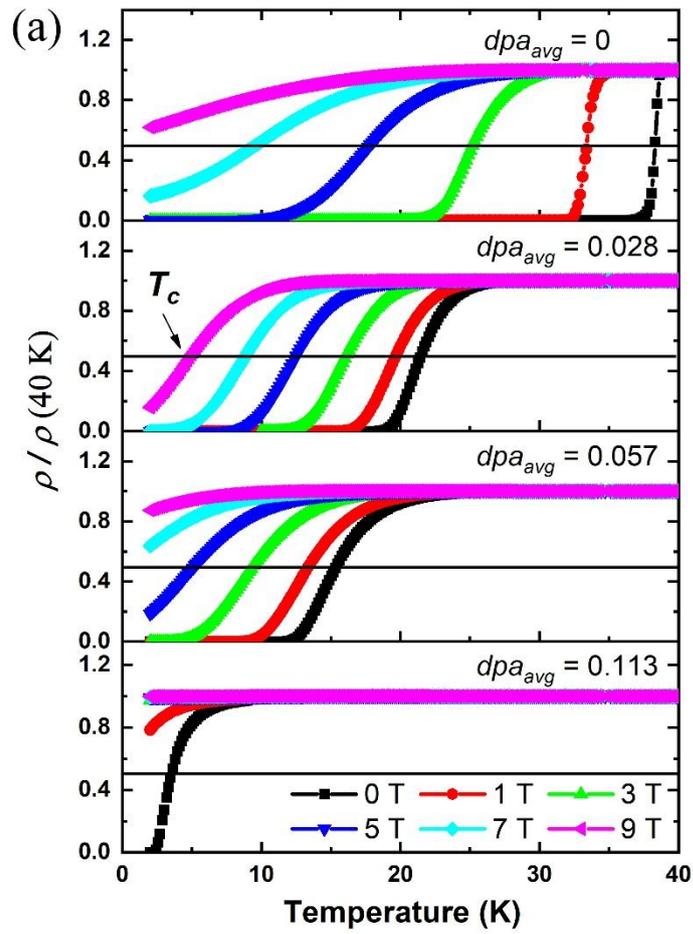

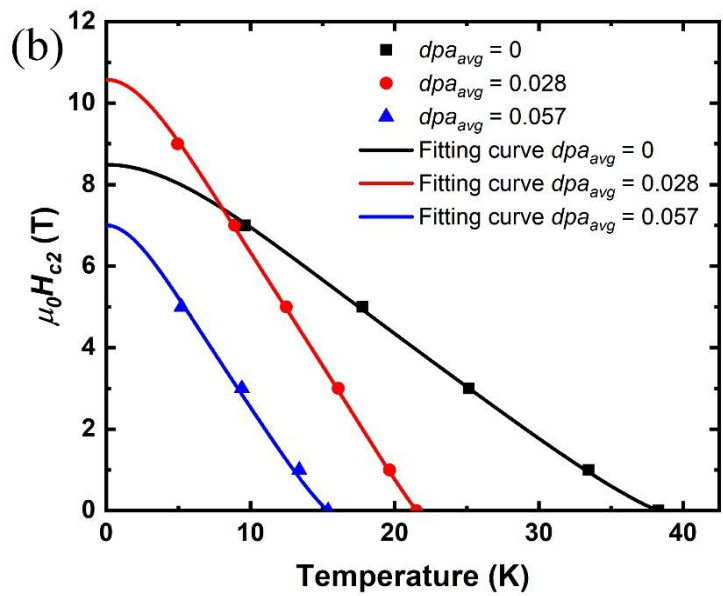

Fig. 4

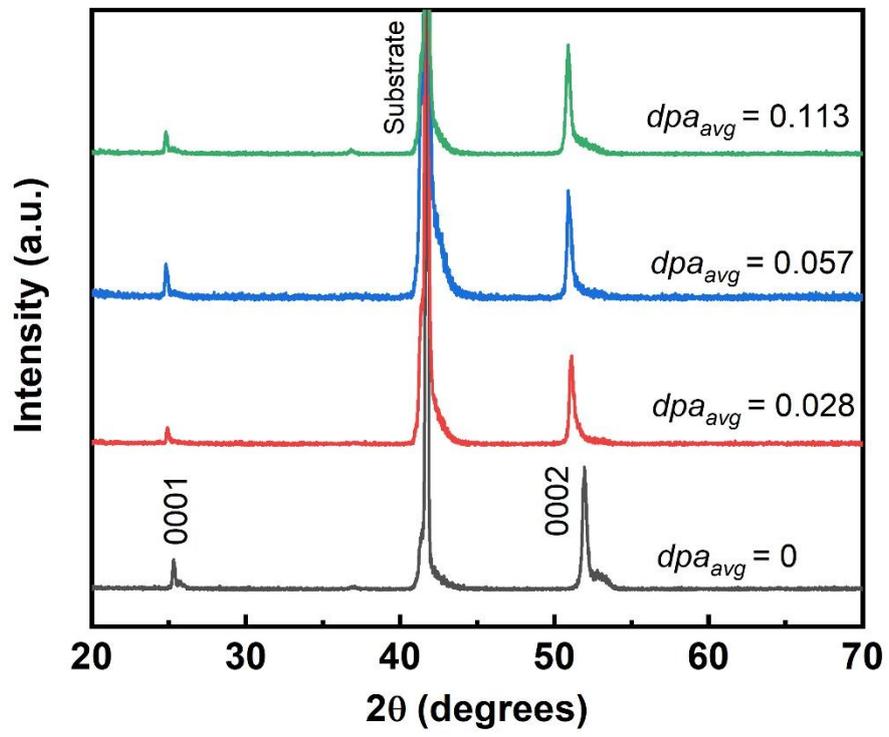

Fig. 5

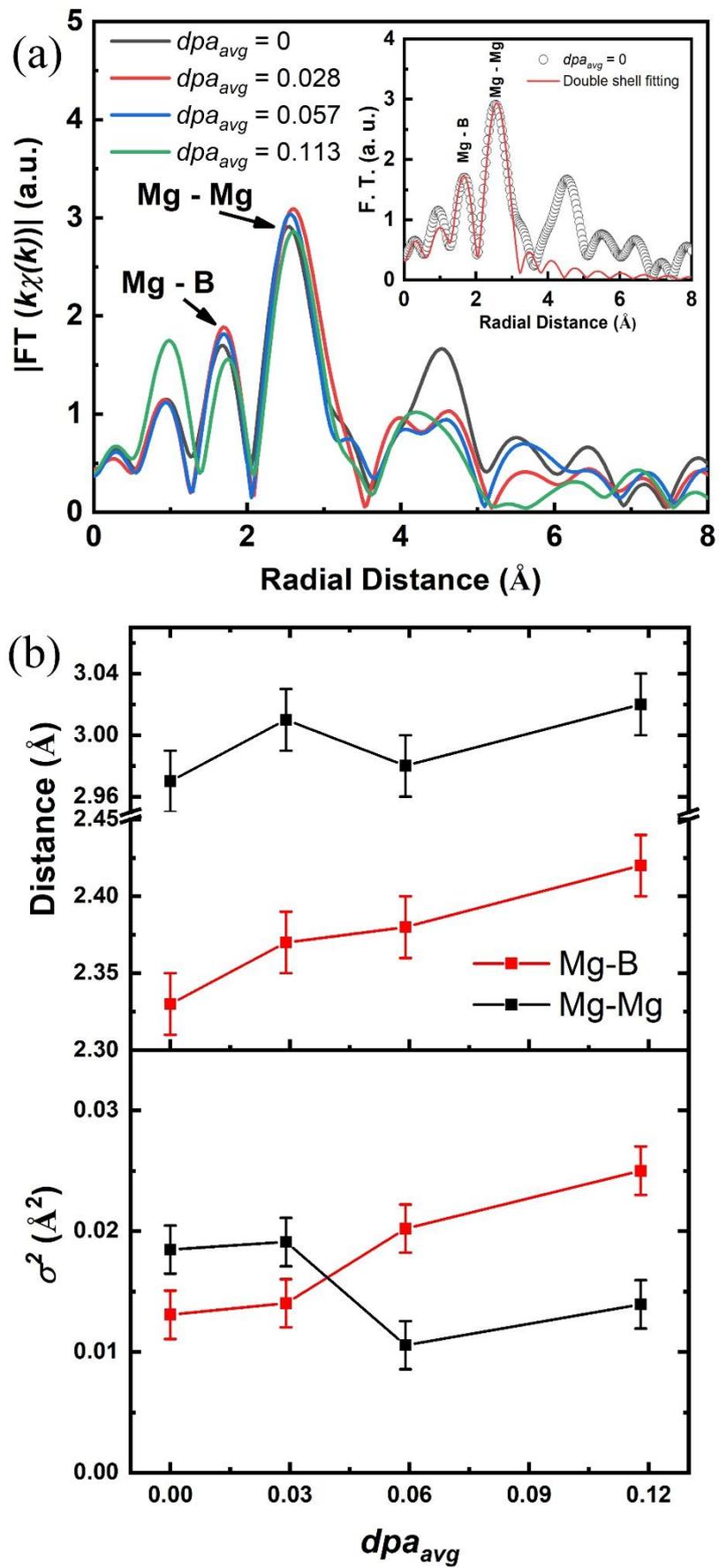

Fig. 6

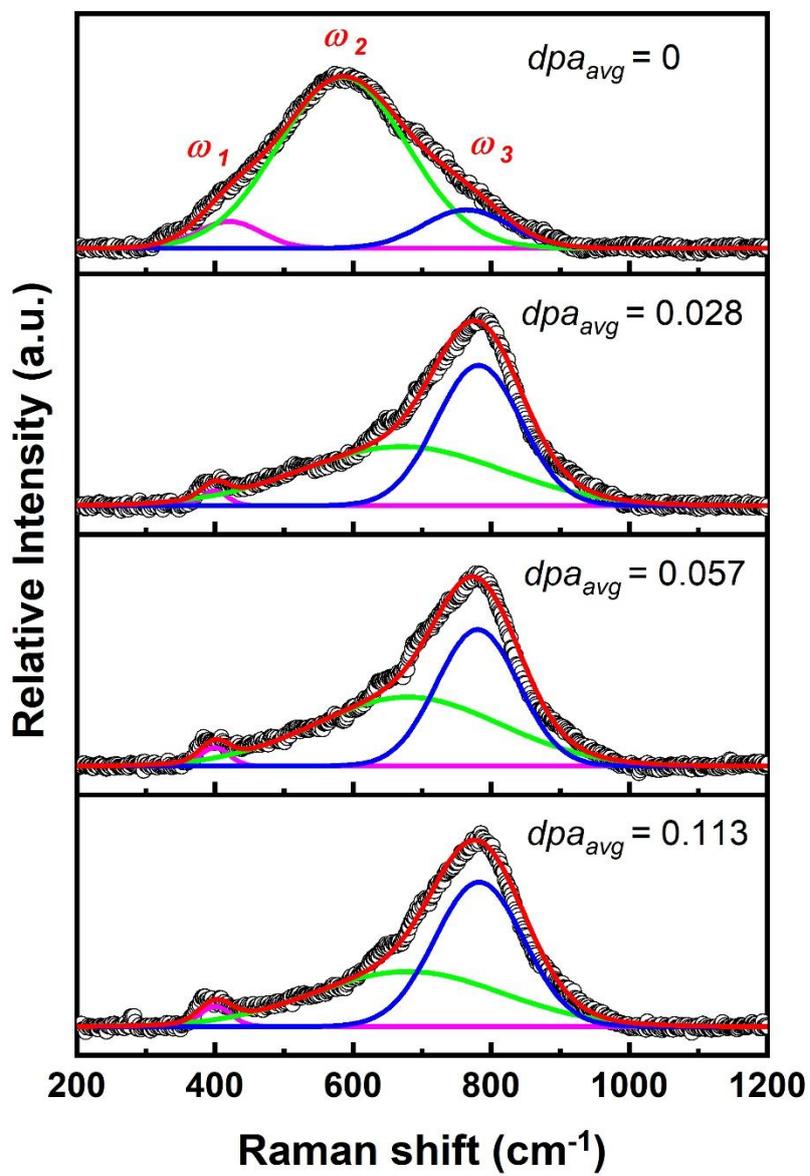

Fig. 7

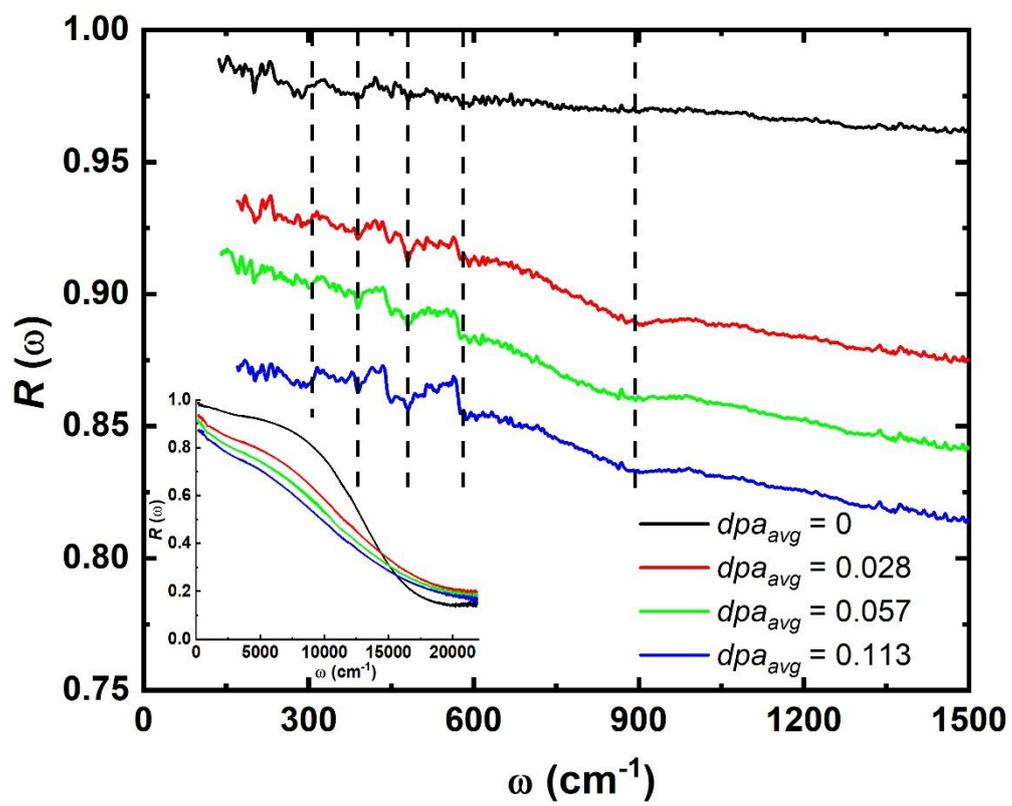

Fig. 8

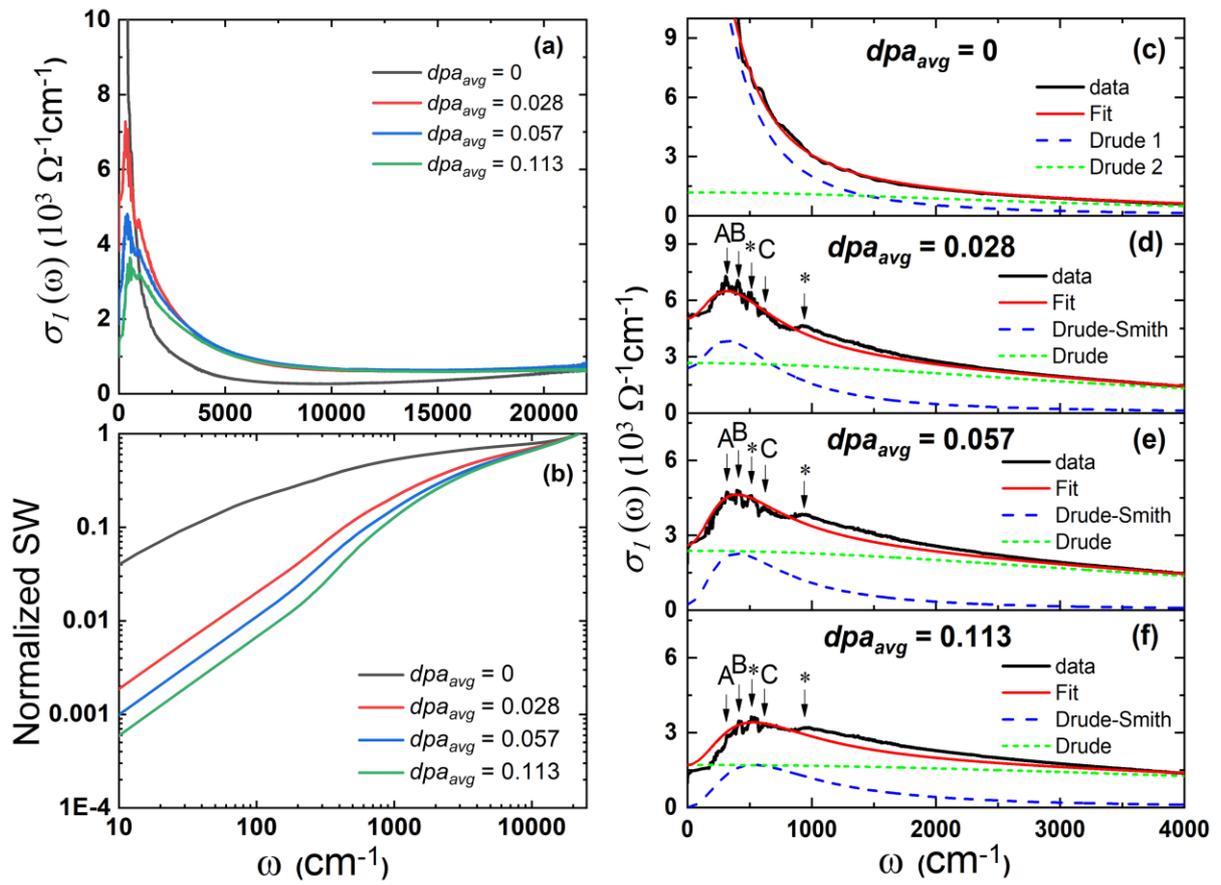

Table 1

| Sample | $c$ - lattice constant (Å) | $\rho(40\ \text{K})$ ($\mu\Omega$ cm) | $T_c$ (K) | $\mu_0 H_{c2}(0\ \text{K})$ (T) | $\omega_1$ (cm$^{-1}$) | $\omega_2$ (cm$^{-1}$) | $\omega_3$ (cm$^{-1}$) | $\lambda$ | $\Theta_D$ (K) |
|---|---|---|---|---|---|---|---|---|---|
| $dpa_{avg} = 0$ | 3.516 | 28.89 | 38.33 | 8.49 | 420 | 585 | 764 | 1.03 | 697 |
| $dpa_{avg} = 0.028$ | 3.569 | 181.84 | 20.97 | 10.57 | 400 | 671 | 782 | 0.74 | 740 |
| $dpa_{avg} = 0.057$ | 3.580 | 271.79 | 14.57 | 6.99 | 400 | 679 | 761 | 0.65 | 744 |
| $dpa_{avg} = 0.113$ | 3.584 | 373.47 | 3.02 | N/A | 401 | 680 | 783 | 0.45 | 745 |

* N/A: not available

Table 2

| Sample | $\omega_{p,DS}$ (cm$^{-1}$) | $1/\tau_{imp,DS}$ (cm$^{-1}$) | $c_1$ | $\omega_{p,D}$ (cm$^{-1}$) | $1/\tau_{imp,D}$ (cm$^{-1}$) |
|---|---|---|---|---|---|
| $dpa_{avg}= 0$ | 20126 | 328 | 0 | 15445 | 3356 |
| $dpa_{avg}= 0.028$ | 13450 | 407 | -0.68 | 25126 | 3950 |
| $dpa_{avg}= 0.057$ | 10553 | 410 | -0.95 | 25850 | 4673 |
| $dpa_{avg}= 0.113$ | 10459 | 532 | -1 | 26291 | 6737 |